\documentclass[conference]{IEEEtran}
\normalsize
\ifCLASSINFOpdf
\else
\fi

\usepackage{amsthm}
\usepackage{amsmath}
\usepackage{amssymb}
\usepackage{graphicx}
\usepackage{esint}
\usepackage{amsfonts}
\usepackage{cite}
\usepackage{balance}
\usepackage{caption}
\usepackage{subcaption}
\usepackage{epstopdf}
\usepackage{color}
\usepackage{tikz}
\usepackage{pgfplots} 
\usetikzlibrary{patterns}
\makeatletter

\theoremstyle{plain}
\newtheorem{thm}{\protect\theoremname}

\makeatother

\providecommand{\theoremname}{Theorem}

\theoremstyle{plain}

\makeatother

\providecommand{\lemmaname}{Lemma}

\begin{document}
\title{On the Ergodic Secrecy Capacity of Reconfigurable Intelligent Surface Aided Wireless Systems Under Mixture Gamma Fading}

\author{\IEEEauthorblockN{Alexandros-Apostolos A. Boulogeorgos\IEEEauthorrefmark{1}\IEEEauthorrefmark{2}, and Angeliki Alexiou\IEEEauthorrefmark{1}}
	
		\IEEEauthorrefmark{1}Department of Digital Systems, University of Piraeus, Piraeus 18534,  Greece, \\
		E-mails: al.boulogeorgos@ieee.org, alexiou@unipi.gr \\
		\IEEEauthorrefmark{2} Department of Electrical and Computer Engineering, University of Western Macedonia, Kozani, Greece\\
}
\maketitle	

\begin{abstract}
This paper presents a quantified assessment of the physical layer security capabilities of reconfigurable intelligent surface (RIS)-aided wireless systems under eavesdropping. Specifically, we derive a closed-form expression for the ergodic secrecy capacity (ESC) that is adaptable to different types of fading and RIS size. The channels between the transmitter (TX) and RIS, the RIS and legitimate receiver as well as the TX and eavesdropper are assumed to follow independent mixture Gamma (MG) distributions. Note that  MG is capable of modeling a large variety of well-known distributions, including Gaussian, Rayleigh, Nakagami-$m$, Rice, and others. The results reveal that as the RIS size increases, although the legitimate links diversity order increases, the ESC gain~decreases. 
\end{abstract}


\section{Introduction}\label{S:Intro}
\IEEEPARstart{R}{econfigurable} intelligent surfaces (RISs) allow flexible manipulation of the propagation environment creating additional degrees of freedom, which in turn enables enhanced physical layer security (PLS) capabilities~\cite{Boulogeorgos2021,Wan2021,Tsiftsis2022,Ntontin2021,Boulogeorgos2022a,Ntontin2022,Boulogeorgos2020a,Zhang2022,Renzo2020,Tang2021,Danufane2021}. Inspired by this, a great amount of effort has been put on designing RIS phase shift (PS) selection strategies that maximizes the system's  secrecy capacity and theoretical frameworks that quantify the PLS capabilities of RIS-aided wireless~systems. 

From the RIS PS selection strategy design perception, the authors of~\cite{Cui2019} formulated and solved a secrecy capacity maximization problem through jointly designing the access point transmit and RIS reflect beam vectors in a RIS-aided secure wireless  system, where a multi-antenna access point sends confidential messages to a single-antenna user in the presence of a single-antenna eavesdropper. In~\cite{Chu2020}, the authors presented a joint power allocation and RIS PS design strategy that minimizes the transmit power subject to the secrecy capacity constraint at the legitimate user in RIS-aided wireless systems. In~\cite{Shen2019}, a  secrecy capacity maximization policy that accounts for the source transmit power and PS constraints at the RIS was documented. In~\cite{Hong2021}, the  design of robust, secure, and energy efficient transmission in RIS-aided wireless systems by jointly optimizing the transmission beam and RIS PS vectors as well as the artificial noise covariance matrix  was reported. Finally, the authors of~\cite{Sun2022} studied the problem of sum rate maximization without violating the PLS requirements of a RIS-aided multi-user cellular system by jointly designing the receive decoder at the end-users, the digital precoder and the artificial noise at the base station, as well as the PS at the~RIS.

From the performance analysis point of view, in~\cite{Yang2020}, the authors investigated the secrecy outage performance of a RIS-aided wireless system, in which both the legitimate user and the eavesdropper receive the information signal via the same RIS.  In~\cite{Xu2021}, an approximation of ergodic secrecy rate of RIS-aided wireless systems in the presence of eavesdropper was documented. The authors of~\cite{Lv2021} quantified the secrecy capacity performance of two-way RIS-aided wireless systems, assuming that all the established channels can be modeled as Rayleigh RVs. In~\cite{Trigui2021}, the authors reported a theoretical framework for the evaluation of secrecy outage probability and ergodic secrecy capacity in the presence of discrete phase noise. 
Finally, in~\cite{Ai2021}, the authors quantified the secrecy outage performance of RIS-aided vehicle-to-vehicle and vehicle-to-infrastructure wireless systems, in which the eavesdropper establishes a direct link with the source, while no link between the RIS and the eavesdropper exits.  
All the aforementioned works assumed that all the established channels can be modeled as Rayleigh distributed random variables (RVs).

The secrecy performance of RIS-aided wireless systems are highly dependent on the  statistics of the established communication channels. However, to the best of the authors knowledge, all the so far published contributions  assumed that the established links experience Rayleigh fading. A general theoretical framework that accounts for different types of channels could serve as a useful tool for quantification of PLS capabilities of RIS-aided wireless systems in different propagation environments. Motivated by this,  this contribution focuses on the investigation of the secrecy performance of such systems under different channel conditions. In more detail, we present a comprehensive system model in which the transmitter (TX) communicates with the legitimate user via an RIS, while the eavesdropper directly receives the information signal from the receiver.  In contrast to previous publications, we assume mixture Gamma (MG) fading in both the TX to RIS and RIS to legitimate receiver (RX) as well as TX to eavesdropper channels. Note that MG has been proven to be capable of accurate modeling  an important number of fading conditions including but not limited to Rayleigh, Rice, Nakagami-$m$, Gamma, and generalized Gamma~\cite{Boulogeorgos2022}. Building upon the aforementioned system model, we extract a novel theoretical framework for the quantification of the ergodic secrecy capacity of RIS-aided wireless system. This framework evaluates the PLS performance envelop and provides useful insights for the design of PLS in RIS-aided wireless systems.

\subsubsection*{Notations} 
Unless otherwise stated, the operators $\mathbb{E}\left[\cdot\right]$, $\exp\left(x\right)$, $\log_2\left(x\right)$ and $\ln(x)$
denote the statistical expectation, the  exponential function, the base-$2$ logarithm of x, and the natural logarithm of $x$, respectively. Moreover, $\mathrm{K}_n(x)$ stands for the modified Bessel function of the second kind and of the $n-$th order~\cite[Eq. (8.407)]{B:Gra_Ryz_Book}.
The upper incomplete Gamma functions \cite[Eq. (8.350/2)]{B:Gra_Ryz_Book} is represented by  $\Gamma\left(\cdot,\cdot\right)$, respectively, while the Gamma function \cite[Eq. (8.310)]{B:Gra_Ryz_Book} is denoted by $\Gamma\left(\cdot\right)$. 
Finally, $\,_2F_1\left(\cdot,\cdot;\cdot;\cdot\right)$ stands for the Gauss hypergeomentric function~\cite[Eq. (4.1.1)]{B:Abramowitz}, while $\mathrm{G}_{p,q}^{m,n}\left(x\left|\begin{array}{c} a_1, a_2,\cdots,a_p\\b_1, b_2,\cdots, b_q\end{array}\right.\right)$ denotes the Meijer's G-function~\cite[Eq. (9.301)]{B:Gra_Ryz_Book}.

\section{System and signal model}\label{sec:SSM}

\begin{figure}
	\centering
	\scalebox{0.5}{\input{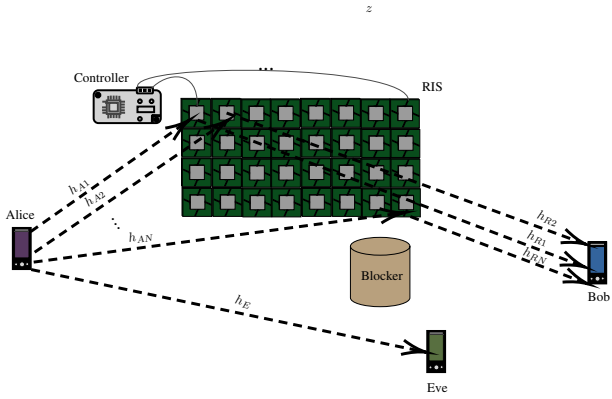}}
	\caption{System model.}
	\label{Fig:SM}
\end{figure}

As illustrated in Fig.~\ref{Fig:SM}, we consider a RIS-aided wireless system that consists of a TX, a legitimate RX and an eavesdropper. In what follows, we refer to the TX as Alice, the legitimate RX as Bob and the eavesdropper as Eve. Alice, Bob and Eve are assumed to be equipped by a single antenna. The RIS has $N$ meta-atoms (MAs) that are coordinated by a controller. No direct link can be established between Alice and Bod. Thus, Alice communicates with Bob through the RIS. The baseband equivalent received signal at Bob can be expressed~as
\begin{align}
	r_B = \beta_B A\, s + n_B,
	\label{Eq:r_B}
\end{align}
where $s$ is Alice transmission symbol, $n_B$ is a zero-mean Gaussian distributed RV with variance equal to $N_o$ that models the additive white Gaussian noise (AWGN) and $\beta_B$ stands for the end-to-end geometric gain, which depends on the Bob's and Alice antenna gains, the RIS gain, the Bob-RIS and RIS-Alice distance, and the transmission frequency.
Moreover, $A$ denotes the Alice-RIS-Bob end-to-end-channel, which according to~\cite{Boulogeorgos2020}, can be evaluated~as
\begin{align}
	A = \sum_{i=1}^{N} h_{A,i}\,g_i\, h_{R,i},
	\label{Eq:A}
\end{align}
where $h_{A,i}$ and $h_{R,i}$ are respectively the Alice-$i-$th MA and $i-$th MA-Bob channel coefficients. Likewise, $g_i$ stands for the $i-$th MA response and can be obtained~as
\begin{align}
	g_i = \left|g_i\right|\,\exp\left(j\phi_{g_i}\right),
\end{align}
with  $ \left|g_i\right|$ and $\phi_{g_i}$ being respectively the amplitude and phase of $g_i$. 
Let $\phi_{h_{A,i}}$ and $\phi_{h_{R,i}}$ respectively be the $h_{A,i}$ and $h_{R,i}$ phases. By assuming that the RIS controller has fully knowledge of $\phi_{h_{A,i}}$ and $\phi_{h_{R,i}}$  for all $i\in[1, N]$, the optimal phase response of the $i-$th MA can be obtained~as
\begin{align}
	\phi_{g_i} = - \phi_{h_{A,i}} - \phi_{h_{R,i}}.
	\label{Eq:phi_g_i}
\end{align}
Without loss of generality, $\left|g_i\right|=1$ for all $i\in[1, N]$. Thus, by applying~\eqref{Eq:phi_g_i} to~\eqref{Eq:A}, the Alice-RIS-Bob end-to-end channel can be expressed~as
\begin{align}
	A = \sum_{i=1}^{N}\left|h_{A,i}\right|\, \left|h_{R,i}\right|. 
	\label{Eq:A_step_1}
\end{align}
Note that $\left|h_{A,i}\right|$ and $\left|h_{R,i}\right|$ are independent MG RVs with probability density functions (PDFs) that can be respectively expressed~as
\begin{align}
	f_{h_{A,i}}\left(x\right) = \sum_{m=1}^{M} 2\,a_{m}^{(1)}\, x^{2\,b_m^{(1)}-1} \, \exp\left(-c_1\,x^2\right) 
	\label{Eq:f_h_Ai}
\end{align}
and 
\begin{align}
	f_{h_{R,i}}\left(x\right) = \sum_{k=1}^{K} 2\,a_{k}^{(2)}\, x^{2\,b_k^{(2)}-1} \, \exp\left(-c_2\,x^2\right), 
	\label{Eq:f_h_Ri}
\end{align}
where $M$ and $K$ represents the numbers of terms that have been used to approximate the PDF of $\left|h_{A,i}\right|$ and  $\left|h_{R,i}\right|$, respectively. Moreover,  $a_{m}^{(1)}$ and $b_m^{(1)}$ with $m\in[1, M]$ and $c_1$ are parameters of the $m-$th term of~\eqref{Eq:f_h_Ai}. Finally,  $a_{k}^{(2)}$ and $b_k^{(2)}$ with $k\in[1, K]$ and $c_2$ are parameters of the $m-$th term of~\eqref{Eq:f_h_Ri}.

By assuming that a direct link can be established between Alice and Eve and no-direct link exists between the RIS and Eve\footnote{Note that this is considered a realistic assumption, since the RIS steer the incident beam towards Bob and not towards Eve; thus, Eve captures a very small fraction of the reflected power from a secondary RIS lobe, which is expected to be below the noise threshold. As a result, Eve can only receive signal directly from Alice~\cite{Boulogeorgos2021a,Dai2020,Liu2021,Lin2021}.}, the received baseband equivalent at Eve can be obtained~as
\begin{align}
	r_{E} = \beta_E\, h_{E}\, s + n_{E}, 
	\label{Eq:r_E}
\end{align}
where $n_{E}$ is a zero-mean Gaussian RV with variance $N_o$ that models the AWGN. Additionally, $\beta_E$ is the geometric gain of the Alice-Eve link. Moreover, $h_E$ denotes the Alice-Eve channel fading coefficient. The envelop of $h_E$ is modeled as a MG RV with PDF that can be written~as
\begin{align}
	f_{h_{E}}\left(x\right) = \sum_{l=1}^{L} 2\,a_{l}^{(3)}\, x^{2\,b_l^{(3)}-1} \, \exp\left(-c_3\,x^2\right), 
	\label{Eq:f_h_E}
\end{align}
where $L$ denotes the number the numbers of terms  employed to approximate the PDF of $\left|h_E\right|$. In addition, $a_{l}^{(3)}$, $b_l^{(3)}$ with $l\in[1, L]$ and $c_3$ are parameters of the $l-$th term of~\eqref{Eq:f_h_E}. 

\section{Performance analysis}
By assuming that the TX has only average channel state information (CSI) and no instantaneous CSI of either the legitimate nor  eavesdropping links, the secrecy capacity is defined~as
\begin{align}
	C_{s} = C_{B}-C_{E}.
	\label{Eq:Cs_instance}
\end{align}
where, $C_B$ stands for the legitimate channel capacity that can be expressed~as
\begin{align}
	C_B = \log_{2}\left(1+\gamma_B\right),
	\label{Eq:C_B}
\end{align}
with $\gamma_B$ being the instantaneous signal-to-noise-ratio (SNR) at Bob.
Notice, that, according to~\eqref{Eq:Cs_instance}, the instantaneous secrecy capacity can take both negative and positive values. If the instantaneous secrecy capacity is positive, by selecting a suitable transmission data rate, physical layer security can be achieved. On the contrary, if $C_s\leq 0$, physical layer security cannot be achieved. 
    
From~\eqref{Eq:r_B}, the instantaneous SNR at Bob can be obtained~as
\begin{align}
	\gamma_{B} = \frac{\beta_B^2\,A^2\,P_s}{N_o},
	\label{Eq:gamma_B}
\end{align}
where $P_s$ is Alice average transmission power. 
Similarly, $C_E$ stands for the eavesdropping channel capacity, which can be analyzed~as 
\begin{align}
	C_E = \log_{2}\left(1+\gamma_E\right),
	\label{Eq:C_E}
\end{align}
where $\gamma_E$ denotes the SNR at Eve.
From~\eqref{Eq:r_E}, the instantaneous SNR at Eve can be expressed~as
\begin{align}
	\gamma_{E}=\frac{\beta_E^2\,\left|h_{E}\right|^2\,P_s}{N_o}. 
	\label{Eq:gamma_E}
\end{align}

The following theorem returns a closed-form expressions for the ergodic secrecy capacity.
\begin{thm}
	The ergodic secrecy capacity can be obtained~as in~\eqref{Eq:C_s_final}, given at the top of the next page. 
	\begin{figure*}
		\begin{align}
			C_s &=\frac{1}{\ln(2)}\left(\frac{\beta_B^2\,P_s}{ N_o}\right)^{-\frac{k_A+m_A}{2}}  \frac{ \Xi^{k_A + m_A}}{\Gamma\left(k_A\right)\Gamma\left(m_A\right)}
			 \mathrm{G}_{2,4}^{4,1}\left(\frac{\Xi^2\,N_o}{\beta_B^2\,P_s}\left|\begin{array}{c} -\frac{k_A+m_A}{2}, 1-\frac{k_A+m_A}{2}\\ \frac{k_A-m_A}{2}, - \frac{k_A-m_A}{2}, -\frac{k_A+m_A}{2}, -\frac{k_A+m_A}{2}\end{array}\right.\right)
			\nonumber \\ & 
			- \sum_{l=1}^{L} \frac{ a_{l}^{(3)}\, c_3^{-b_l^{(3)}}}{\ln(2)} 
			\mathrm{G}_{4,3}^{1,4}\left(\frac{\beta_{E}^{2}\,P_s}{c_3\, N_o}\left|\begin{array}{c} 0, 0, 1-b_l^{(3)}, -b_l^{(3)}\\ 0, -b_l^{(3)}, -1 \end{array}\right.\right)
			\label{Eq:C_s_final}
		\end{align} 
		\hrulefill
	\end{figure*}
In~\eqref{Eq:C_s_final}, 
\begin{align}
	k_A = - \frac{b_A}{2 a_A} + \frac{\sqrt{b_A^2 - 4 a_A c_A}}{2 a_A},
	\label{Eq:k_A}
\end{align}  
\begin{align}
	m_A = - \frac{b_A}{2 a_A} - \frac{\sqrt{b_A^2 - 4 a_A c_A}}{2 a_A}
\end{align}
and 
\begin{align}
	\Xi = \sqrt{\frac{k_A m_A}{\Omega_A}}.
	\label{Eq:Xi_A}
\end{align}
In~\eqref{Eq:k_A}--\eqref{Eq:Xi_A},
\begin{align}
	a_A \hspace{-0.15cm}&=\hspace{-0.15cm} \mu_{A}\left(6\right) \mu_{A}\left(2\right) + \left(\mu_{A}\left(2\right)\right)^2 \mu_{A}\left(4\right) - 2\left(\mu_{A}\left(4\right)\right)^2,
	\\
	b_A\hspace{-0.15cm}&=\hspace{-0.15cm} \mu_{A}\left(6\right) \mu_{A}\left(2\right)- 4 \left(\mu_{A}\left(4\right)\right)^2 + 3 \left(\mu_{A}\left(2\right)\right)^2 \mu_{A}\left(4\right),
	\\
	c_A &= 2 \left(\mu_{A}\left(2\right)\right)^2 \mu_{A}\left(4\right)
\end{align}
and 
\begin{align}
	\Omega_A = \mu_A\left(2\right),
\end{align}
where 
\begin{align}
	\mu_{A}\left(l\right) = \sum_{l_1=0}^{l}&\sum_{l_2=0}^{l_1}\cdots\sum_{l_{N-1}=0}^{l_{N-2}}
	\left(\begin{array}{c}l\\l_1\end{array}\right) \left(\begin{array}{c}l_1\\l_2\end{array}\right)
	\cdots 
	\left(\begin{array}{c}l_{N-2}\\l_{N-1}\end{array}\right)
	\nonumber \\ 
	& \hspace{-0.7cm} \times \mu_{\chi_1}\left(l-l_1\right) \mu_{\chi_2}\left(l_1-l_2\right) 
	\cdots \mu_{\chi_{N-1}}\left(l_{N-1}\right)
	\label{Eq:mu_A}
\end{align}
and 
\begin{align}
	\mu_{\chi_i}(l) = \sum_{m=1}^{M}&\sum_{k=1}^{K}
	a_m^{(1)} a_k^{(2)} \left(\frac{c_1}{c_2}\right)^{-\frac{b_m^{(1)}-b_k^{(2)}}{2}} \left(c_1 c_2\right)^{-\frac{b_m^{(1)}+b_k^{(2)}+n}{2}}
	\nonumber \\ & \times
	\Gamma\left(b_{m}^{(1)}+\frac{n}{2}\right) \Gamma\left(b_{k}^{(2)}+\frac{n}{2}\right)
	\label{Eq:mu_x}
\end{align}
\end{thm}
\begin{IEEEproof}
	For brevity, the proof is provided in the Appendix.
\end{IEEEproof}

\section{Results \& Discussion}

This section focuses on verifing the theoretical framework by means of Monte Carlo simulations. The following scenario is considered.  It is assumed that the Alice-RIS and RIS-Bob channel coefficients follow independent and identical Rice distributions with $K_r$ parameter equal to $5\,\mathrm{dB}$, while the Alice-Eve channel coefficient is modeled as a Nakagami-$m$ RV with $m=2$. For an accurate approximation of the Rice distribution, we select $M=K=20$, 
\begin{align}
	a_{n}^{(i)} = \frac{\delta\left(K_r, n\right)}{\sum_{k_1=1}^{N_r} \delta\left(K_r, k_1\right) \Gamma\left(b_{k_1}^{(2)}\right) c_i^{-b_{k_1}^{(i)}}},
	\label{Eq:a_k_rice}
\end{align}    
where
	$\delta\left(K_r, n\right) = \frac{K_r^{n-1}\left(1+K_r\right)^n}{\exp\left(K_1\right) \left( (n-1)!\right)^2},$
while $c_i=1+K_r$, and $b_{n}^{(i)}=n$. Note that $i\in\{1, 2\}$ and
\begin{align}
	N_r=\left\{\begin{array}{c}M, \, i=1 \\ K, \, i=2 \end{array}\right..
\end{align}
The MG distribution can be simplified to Nakagami-$m$ by setting $L=1$, $a_l^{(3)}=\frac{m^{m}}{\Gamma(m)}$, and $b_l^{(3)}=c_3=m$ with $m$ being the shape parameter. 

\begin{figure}
	\centering
	\scalebox{1.00}{\begin{tikzpicture}
	\begin{axis}[
		legend columns=1,
		legend entries={$\beta_{E}^{2}=-5\,\mathrm{dB}$, $\beta_{E}^{2}=0\,\mathrm{dB}$, $\beta_{E}^{2}=5\,\mathrm{dB}$},
		ybar legend,
		ymajorgrids=true,
		legend pos = north west,
		xlabel = {$N$},
		xtick={4, 8, 12, 16, 20, 24, 28},
		ylabel ={$C_s\,\mathrm{(bits/s/Hz)}$},
		]
		
		\addplot[ybar, pattern=north east lines] coordinates {
			(4.0, 3.34272) 
			(6.0, 4.507684643969879)
			(8.0, 5.327193755602597)
			(10.0, 5.780307297325157)
			(12.0, 6.257356290460929)
			(14.0, 6.68391028212764)
			(16.0, 7.066345365356715)
			(18.0, 7.424168116566778)
			(20.0, 7.732992038083899)
			(22.0, 8.01434235305475)
			(24.0, 8.272335859659929)
			(26.0, 8.51032281490735)
			(28.0, 8.731034537612176)
			(30.0, 8.936710677446557)
		};
	\addplot[ybar, pattern=horizontal lines] coordinates {
		(4.0, 2.80787) 
		(6.0, 3.9639555649364207)
		(8.0, 4.78346)
		(10.0, 5.236578218291699)
		(12.0, 5.713627211427471)
		(14.0, 6.1401812030941825)
		(16.0, 6.522616286323257)
		(18.0, 6.88043903753332)
		(20.0, 7.189262959050441)
		(22.0, 7.470613274021294)
		(24.0, 7.728606780626472)
		(26.0, 7.966593735873894)
		(28.0, 8.18730545857872)
		(30.0, 8.392981598413101)
	};
		\addplot[ybar, pattern=horizontal lines, color=black] coordinates {
		(4.0, 1.77161) 
		(6.0, 3.017562022568158)
		(8.0, 3.837071134200876)
		(10.0, 4.290184675923436)
		(12.0, 4.767233669059208)
		(14.0, 5.19378766072592)
		(16.0, 5.576222743954994)
		(18.0, 5.934045495165058)
		(20.0, 6.242869416682178)
		(22.0, 6.524219731653031)
		(24.0, 6.782213238258209)
		(26.0, 7.020200193505631)
		(28.0, 7.2409119162104565)
		(30.0, 7.446588056044837)
	};

	\end{axis}
\end{tikzpicture}}
	\vspace{-0.25cm}
	\caption{$C_s$ vs $N$ for different values of $\beta_E^2$, assuming $\beta_B^2=0\,\mathrm{dB}$.}
	\label{Fig:BarFig}
\end{figure}
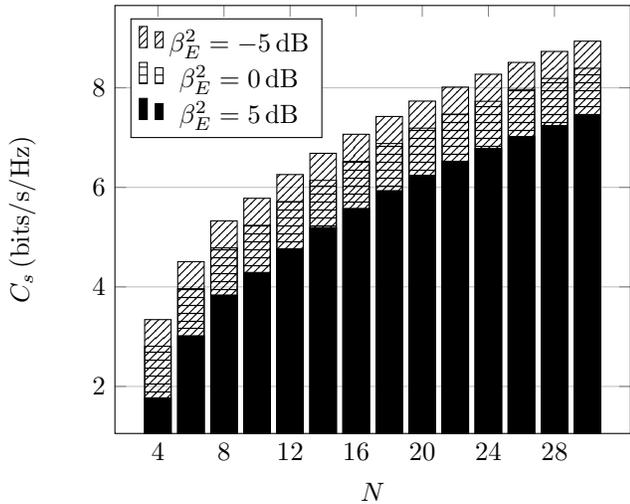 
Figure~\ref{Fig:BarFig} depicts the ergodic secrecy capacity as a function of $N$ for different values of $\beta_{E}^2$, assuming that $\beta_B^{2}=0$. As expected, for a given $N$, as the eavesdropping channel improves, i.e. as $\beta_{E}^{2}$ increases, the ergodic secrecy capacity decreases. For example, for $N=4$, the ergodic secrecy capacity decreases from  $3.34$ to $2.81\,\mathrm{bits/s/Hz}$, as $\beta_{E}^{2}$ increases from $-5$ to $0\,\mathrm{dB}$. In other words, a $15.9\%$ ergodic secrecy capacity degradation is observed. Similarly, for $N=8$,  the ergodic secrecy capacity decreases from $5.33$ to $4.78\,\mathrm{bits/s/Hz}$, for the same $\beta_{E}^{2}$  increase. This is translated into a $10.2\%$ ergodic secrecy capacity degradation. From these examples, it becomes evident that as $N$ increases, the ergodic secreacy capacity degradation, due to eavesdropping channel improvement, decreases. Moreover, from this figure, it becomes apparent that, for a fixed $\beta_{E}^{2}$, as $N$ increases, the system's diversity order increases; thus, the ergodic secrecy capacity increases. For instance, for $\beta_{E}^{2}=0\,\mathrm{dB}$, the ergodic secrecy capacity increases from $2.81$ to $3.96\,\mathrm{bits/s/Hz}$ as $N$ increases from $4$ to $6$, and from  $6.14$ to $6.52\,\mathrm{bits/s/Hz}$, as $N$ increases from $14$ to $16$. This indicates that, although as $N$ increases, the diversity order increases, the diversity gain to the ergodic secrecy capacity decreases. 

\begin{figure}
	\centering
	\scalebox{1.00}{\input{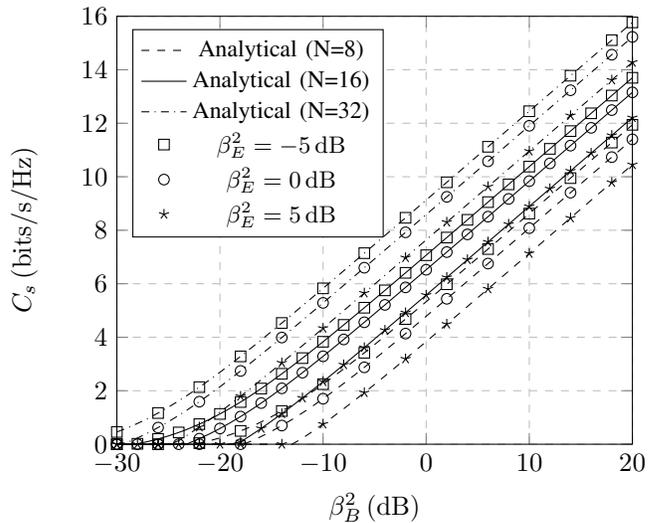}}
	\vspace{-0.25cm}
	\caption{$C_s$ vs $\beta_B^2$, for different values of $\beta_E^2$ and $N$.}
	\label{Fig:PlotFig}
\end{figure} 
Figure~\ref{Fig:PlotFig} illustrates the ergodic secrecy capacity as a function of $\beta_B^2$, for different values of $\beta_E^2$ and $N$. We observe that for given $\beta_{E}^{2}$ and $N$, as $\beta_B^{2}$ increases, the legitimate channel improves; as a result, the ergodic secrecy capacity increases. For example, for $\beta_E^{2}=0\,\mathrm{dB}$  and $N=32$, the ergodic secrecy capacity increases from $5.28$ to $8.59$, as $\beta_B^2$ increases from $-10$ to $0\,\mathrm{dB}$. Moreover, for fixed $\beta_{B}^2$ and $N$, as $\beta_{E}^{2}$ increases, the ergodic secrecy capacity decreases. For instance, for $\beta_B^2=0\,\mathrm{dB}$ and $N=32$, the ergodic secrecy capacity changes from $8.59$ to $7.64\,\mathrm{bits/s/Hz}$, as $\beta_{E}^{2}$ increases from $0$ to $5\,\mathrm{dB}$. Finally, for given $\beta_B^{2}$ and  $\beta_{E}^{2}$, as $N$ increases, the diversity order and gain increase; hence the ergodic secrecy capacity increases.  

 \section{Conclusions}
 
 In this paper, we investigated the PLS capabilities of RIS-aided wireless systems in the presence of eavesdropper. We derived a novel and general closed-form expression for the ergodic secrecy capacity that accounts for both different types of fading and different RIS sizes. The results highlighted that beyond a specific RIS size, i.e., number of RIS elements, a further increase of the RIS size will not result to important PLS capabilities improvements.  
 
 \section*{Appendix}

\section*{Proof of Theorem 1}
 Based on~\eqref{Eq:Cs_instance}, the ergodic secrecy capacity can be evaluated~as
\begin{align}
	\overline{C}_s = \mathbb{E}\left[C_s\right],
\end{align} 
or
\begin{align}
	\overline{C}_s = \mathbb{E}\left[C_{B}-C_{E}\right],
\end{align} 
or equivalently
\begin{align}
		\overline{C}_s = \overline{C}_B- \overline{C}_E,
		\label{Eq:C_s}
\end{align}
where 
\begin{align}
	\overline{C}_B = \mathbb{E}\left[C_{B}\right]
\text{ and }
	\overline{C}_E = \mathbb{E}\left[C_{E}\right].
	\label{Eq:C_E_def}
\end{align}
are respectively the ergodic capacities of the legitimate and eavesdropping channels. 

The ergodic capacity of the legitimate channel can be evaluated~as
\begin{align}
	\overline{C}_B = \int_{0}^{\infty} \log_2\left(1+\frac{\beta_B^2\,P_s}{N_o} x^2\right)\, f_{A}(x)\,\mathrm{d}x,
\end{align}
which, by applying~\cite[Eq.(17)]{Boulogeorgos2022}, can be rewritten~as
\begin{align}
\overline{C}_B = \frac{4 \Xi^{k_A + m_A}}{\Gamma\left(k_A\right)\Gamma\left(m_A\right)} \mathcal{J},
\label{Eq:C_B_step_2}
\end{align}
where
\begin{align}
	\mathcal{J}= \int_{0}^{\infty} x^{k_A+m_A-1} \log_2\left(1+\frac{\beta_B^2\,P_s}{N_o} x^2\right)  \mathrm{K}_{k_A-m_A}\left(2\Xi x\right)\,\mathrm{d}x.
	\label{Eq:J}
\end{align}
To extract a closed-form expression for~\eqref{Eq:J}, we first employ~\cite[Eq. (15.1.1)]{B:Abramowitz}, and rewrite~\eqref{Eq:J}~as
\begin{align}
		\mathcal{J}= \frac{\beta_B^2\,P_s}{\ln(2)\, N_o} \int_{0}^{\infty} & x^{k_A+m_A+1} \,_2F_1\left(1, 1; 2; -\frac{\beta_B^2\,P_s}{N_o} x^2\right) 
		\nonumber \\ & \times 
		\mathrm{K}_{k_A-m_A}\left(2\Xi x\right)\,\mathrm{d}x.
\end{align}
Next, by applying~\cite[Eq. (9.34/7)]{B:Gra_Ryz_Book}, we obtain
\begin{align}
	 \mathcal{J}= \frac{\beta_B^2\,P_s}{\ln(2)\, N_o} \int_{0}^{\infty} & x^{k_A+m_A+1} \, \mathrm{G}_{2, 2}^{1, 2}\left( \frac{\beta_B^2\,P_s}{N_o}\,x^2\left|\begin{array}{c} 0, 0 \\ 0, -1\end{array}\right.\right)
	\nonumber \\ & \times 
	\mathrm{K}_{k_A-m_A}\left(2\Xi x\right)\,\mathrm{d}x,
\end{align}
which, with the aid of~\cite{Wofram:BesselK}, can be rewritten~as
\begin{align}
	\mathcal{J}= \frac{\beta_B^2\,P_s}{2\ln(2)\, N_o} &\int_{0}^{\infty}  x^{k_A+m_A+1} \, \mathrm{G}_{2, 2}^{1, 2}\left( \frac{\beta_B^2\,P_s}{N_o}\,x^2\left|\begin{array}{c} 0, 0 \\ 0, -1\end{array}\right.\right)
	\nonumber \\ & \times 
	\mathrm{G}_{0,2}^{2,0}\left({\Xi^2\,x^2}\left|\begin{array}{c} \frac{k_A-m_A}{2}, - \frac{k_A-m_A}{2}\end{array}\right.\right)
	\,\mathrm{d}x.
	\label{Eq:J_step3}
\end{align}
By setting~$y=x^2$,~\eqref{Eq:J_step3} can be written~as
\begin{align}
	\mathcal{J}= &\frac{\beta_B^2\,P_s}{4\ln(2)\, N_o} \int_{0}^{\infty}  y^{\frac{1}{2}\left(k_A+m_A\right)} \, \mathrm{G}_{2, 2}^{1, 2}\left( \frac{\beta_B^2\,P_s}{N_o}\,y\left|\begin{array}{c} 0, 0 \\ 0, -1\end{array}\right.\right)
	\nonumber \\ & \times 
	\mathrm{G}_{0,2}^{2,0}\left({\Xi^2}\,y\left|\begin{array}{c} \frac{k_A-m_A}{2}, - \frac{k_A-m_A}{2}\end{array}\right.\right)
	\,\mathrm{d}y.
	\label{Eq:J_step4}
\end{align}
By applying~\cite[ch. 2.3]{Mathaia2010},~\eqref{Eq:J_step4} can be expressed~as
\begin{align}
		&\mathcal{J}= \frac{1}{4\ln(2)}\left(\frac{\beta_B^2\,P_s}{ N_o}\right)^{-\frac{k_A+m_A}{2}}
		\nonumber \\ & \times
		\mathrm{G}_{2,4}^{4,1}\left(\frac{\Xi^2\,N_o}{\beta_B^2\,P_s}\left|\begin{array}{c} -\frac{k_A+m_A}{2}, 1-\frac{k_A+m_A}{2}\\ \frac{k_A-m_A}{2}, - \frac{k_A-m_A}{2}, -\frac{k_A+m_A}{2}, -\frac{k_A+m_A}{2}\end{array}\right.\right).
		\label{Eq:J_step5}
\end{align}
With the aid of~\eqref{Eq:J_step5},~\eqref{Eq:C_B_step_2} can be rewritten~as 
\begin{align}
	&\overline{C}_B =\frac{1}{\ln(2)}\left(\frac{\beta_B^2\,P_s}{ N_o}\right)^{-\frac{k_A+m_A}{2}}  \frac{ \Xi^{k_A + m_A}}{\Gamma\left(k_A\right)\Gamma\left(m_A\right)}
	\nonumber \\ & \times  \mathrm{G}_{2,4}^{4,1}\left(\frac{\Xi^2\,N_o}{\beta_B^2\,P_s}\left|\begin{array}{c} -\frac{k_A+m_A}{2}, 1-\frac{k_A+m_A}{2}\\ \frac{k_A-m_A}{2}, - \frac{k_A-m_A}{2}, -\frac{k_A+m_A}{2}, -\frac{k_A+m_A}{2}\end{array}\right.\right).
	\label{Eq:C_B_final}
\end{align}

The ergodic capacity of the eavesdropping channel can be computed~as 
\begin{align}
	\overline{C}_E = \int_{0}^{\infty} \log_{2}\left(1+\frac{\beta_{E}^2\,P_{s}}{N_o} x^2\right) \, f_{h_{E}}\left(x\right)\,\mathrm{d}x,
\end{align}
which, by applying~\eqref{Eq:f_h_E}, can be rewritten~as
\begin{align}
	\overline{C}_E = \sum_{l=1}^{L} 2\,a_{l}^{(3)}\, \mathcal{K}_l,
	\label{Eq:C_E_step_3}
\end{align}
where 
\begin{align}
	\mathcal{K}_l =  \int_{0}^{\infty} x^{2\,b_l^{(3)}-1} \, \exp\left(-c_3\,x^2\right)\, \log_{2}\left(1+\frac{\beta_{E}^2\,P_{s}}{N_o} x^2\right) \,\mathrm{d}x,
	\label{Eq:K_I}
\end{align}
which can be rewritten~as 
\begin{align}
	\mathcal{K}_l = \frac{1}{\ln(2)} \int_{0}^{\infty} &x^{2\,b_l^{(3)}-1} \, \exp\left(-c_3\,x^2\right)\,
	\nonumber \\ & \times \ln\left(1+\frac{\beta_{E}^2\,P_{s}}{N_o} x^2\right) \,\mathrm{d}x.
	\label{Eq:K_I_step_1}
\end{align}
By setting $z=x^2$,~\eqref{Eq:K_I_step_1} 
 \begin{align}
 	\mathcal{K}_l = \frac{1}{2\ln(2)} \int_{0}^{\infty} &z^{\,b_l^{(3)}-1} \, \exp\left(-c_3\,z\right)\,
 	\nonumber \\ & \times \ln\left(1+\frac{\beta_{E}^2\,P_{s}}{N_o} z\right) \,\mathrm{d}z,
 	\label{Eq:K_I_step_2}
 \end{align}
which, by applying~\cite[Eq. (8.352/2)]{B:Gra_Ryz_Book}, can be rewritten~as 
 \begin{align}
	\mathcal{K}_l = \frac{1}{2\ln(2)} \int_{0}^{\infty} &z^{\,b_l^{(3)}-1} \, \exp\left(-c_3\,z\right)\,
	\nonumber \\ & \times
	\mathrm{G}_{2, 2}^{1, 2}\left( \frac{\beta_E^2\,P_s}{N_o}\,z\left|\begin{array}{c} 0, 0 \\ 0, -1\end{array}\right.\right) \,\mathrm{d}z.
	\label{Eq:K_I_step_3}
\end{align}
Moreover, with the aid of~\cite[Eq. (8.352/2)]{B:Gra_Ryz_Book},~\eqref{Eq:K_I_step_3} can be expressed~as 
 \begin{align}
	\mathcal{K}_l = \frac{1}{2\ln(2)} \int_{0}^{\infty} &z^{\,b_l^{(3)}-1} \, \Gamma\left(1, c_3\,z\right)\,
	\nonumber \\ & \times
	\mathrm{G}_{2, 2}^{1, 2}\left( \frac{\beta_E^2\,P_s}{N_o}\,z\left|\begin{array}{c} 0, 0 \\ 0, -1\end{array}\right.\right) \,\mathrm{d}z,
	\label{Eq:K_I_step_4}
\end{align}
which, by applying~\cite[Eq. (15.1.1)]{B:Abramowitz} and~\cite[Eq. (9.34/7)]{B:Gra_Ryz_Book},~yields 
\begin{align}
	\mathcal{K}_l = \frac{1}{2\ln(2)} \int_{0}^{\infty} &z^{\,b_l^{(3)}-1} \, 
	\mathrm{G}_{1,2}^{2,0}\left(c_3\,z\left|\begin{array}{c} 1\\0, 1\end{array}\right.\right)
	\nonumber \\ & \times
	\mathrm{G}_{2, 2}^{1, 2}\left( \frac{\beta_E^2\,P_s}{N_o}\,z\left|\begin{array}{c} 0, 0 \\ 0, -1\end{array}\right.\right) \,\mathrm{d}z.
	\label{Eq:K_I_step_5}
\end{align}
With the aid of~\cite[ch. 2.3]{Mathaia2010},~\eqref{Eq:K_I_step_5} can be rewritten~as 
\begin{align}
	\mathcal{K}_l = \frac{c_3^{-b_l^{(3)}}}{2\ln(2)} \mathrm{G}_{4,3}^{1,4}\left(\frac{\beta_{E}^{2}\,P_s}{c_3\, N_o}\left|\begin{array}{c} 0, 0, 1-b_l^{(3)}, -b_l^{(3)}\\ 0, -b_l^{(3)}, -1 \end{array}\right.\right)
	\label{Eq:K_I_step_6}
\end{align}
By applying~\eqref{Eq:K_I_step_6} in~\eqref{Eq:C_E_step_3}, we~obtain 
\begin{align}
	\overline{C}_E = \sum_{l=1}^{L} & \frac{ a_{l}^{(3)}\, c_3^{-b_l^{(3)}}}{\ln(2)} 
	\nonumber \\ & \times \mathrm{G}_{4,3}^{1,4}\left(\frac{\beta_{E}^{2}\,P_s}{c_3\, N_o}\left|\begin{array}{c} 0, 0, 1-b_l^{(3)}, -b_l^{(3)}\\ 0, -b_l^{(3)}, -1 \end{array}\right.\right).
	\label{Eq:C_E_step_final}
\end{align}
Finally, with the aid of~\eqref{Eq:C_B_final} and~\eqref{Eq:C_E_step_final},~\eqref{Eq:C_s} can be written as in~\eqref{Eq:C_s_final}. This concludes the proof.

\bibliographystyle{IEEEtran}
\bibliography{IEEEabrv,References}
\balance

\end{document}